\documentstyle[12pt,epsfig]{article} 
\def\qed{\hbox{${\vcenter{\vbox{                          %HOLLOW SQUARE
   \hrule height 0.4pt\hbox{\vrule width 0.4pt height 6pt
   \kern5pt\vrule width 0.4pt}\hrule height 0.4pt}}}$}}

\def\shiftdown#1{#1\llap{\lower.04ex\hbox{#1}}}

\begin{document}

\title{Coupling of vector fields at high energies}  
\renewcommand{\thefootnote}{\fnsymbol{footnote}}
\author{Michael L. Schmid\footnotemark[3]\\
Institut f\"ur Theoretische Physik, Universit\"at T\"ubingen\\
         Auf der Morgenstelle 14, D-72076 T\"ubingen, Germany}
 
\footnotetext[3]{E-mail address: michael.schmid@uni-tuebingen.de}
%\address { Institut f\"ur Theoretische Physik, Universit\"at T\"ubingen,
%         Auf der Morgenstelle 14, D-72076 T\"ubingen, Germany }

\date{\today} 

\maketitle

\begin{abstract}
In general relativity and electrodynamics fields are always
generated from static monopoles (like mass or electric charge) or their
corresponding currents by surrounding them in a spherical configuration.
We investigate a generation of fields from primary fields by a scalar coupling.
The generated secondary fields fulfill the condition of source-freedom and therefore 
cannot occur in a spherical configuration. 
The coupling strength depends on the energies of the primary fields.
In most cases these fields can be approximately considered as dipole fields.  
We discuss two applications of couplings for electromagnetic and gravitational spin-1 fields
and for electric and magnetic fields. 
We calculate for both applications the threshold values of field energy for the maximum
coupling strength. \\
The proposed approach yields to a further step towards an unification of
electromagnetism and gravitation and has important consequences for the discrete symmetries.  
\end{abstract}
PACS 11.30.Er \mbox{Charge conjugation, parity, time reversal, and other discrete
symmetries.}
\section{Introduction}
\label{sec:1}
The classical field theory describes the two long ranging fields of electromagnetism 
and gravitation. 
Some investigations were made to find a common formalism for unification of these
fields and interactions \cite{wey19}, even to find a connection between gravitation and
quantum theory \cite{kl26}, but no convincing conception was found
\cite{pau19}. 
(For a survey see also \cite{rst20}). \\
Extended theories of gravitation based on supersymmetry predict new particles.
These quantum theories of gravitation contain not only the spin-2 graviton, but include also 
spin-0 and spin-1 gravitons (graviscalar and graviphoton) as well as spin-3/2 gravitino
for the description of gravitational interactions \cite{sci88}. \\
Composite models of particles provide an alternative approach for gravitational interactions.
The interaction-particles (bosons) are described as composite states of the
fermions \cite{stu88}. \\
Thus the spin-2 graviton can be considered as a bound state of two spin-1 gravitons \cite{pap65}. 
This bounded state should be resulted by the coupling to itself and should  also generate the
self-interactions, which are described by the nonlinear equations of general
relativity.
This non-linearity results from the fact that the spin-2 graviton couples to
everything and thereby also to itself. \\  
Within the covariant formalism of general relativity, 
the gravitational field can be described as a pure vectorfield in an approximation for weak fields,
where self-interactions must be neglected.
In the case of a non-approximate description of vectorfields the
self-interactions must be removed completely.
Thus unlike the spin-2 gravitons, the spin-1 gravitons as also the photons are no sources of themselves,
for which no self-interactions results. \\
This means that the spin-1 graviton couples to everything but not to itself,
which requires a certain symmetry for the gravitational vectorfield.
This is also shown by linear field equations and the appearance of local conservation laws.
Therefore the gravitational vectorfield  has the property of a sourcefree
field and cannot occur in a spherosymmetrical configuration. \\
The coupling to other (non-gravitational) interactions with the break of the spherical symmetry
should generate spin-1 gravitons. \\  
Therefore, it should be possible to discuss the spin-1 graviton (graviphoton) and the classical 
vector field of gravitation together with the photon and the classical 
electromagnetic field within a common formalism of the linear Maxwell equations.
Although the spin-2 gravitons cannot be described within the linear formalism \cite{wei80},
this represents a further step towards the unification of electromagnetic and gravitational interactions. \\ 
The gravitational fields are described by nonlinear equations in general relativity.
Their nonlinearity structure is caused by the self-interaction.
If we assume that mass and all forms of energy are gravitational charges,
only one kind of gravitational charge can exist.
So there is no mechanism to generate dipole fields like in electrodynamics,
where two kinds of charge are possible. \\ 
We conclude that gravitational dipole fields can exist, but they have to be 
source-free (no kind of charge like mass or electric charge). 
Linear field equations like the homogeneous Maxwell equations describe exactly 
those source-free gravitational fields. \\ 
From a counterpart view between sources and fields (p. 368 E. of \cite{mis73})
the priority of the fields should be preferably considered. 
We can have fields without any source but no source without any field.
The source results from a spherical configuration of the field.
Thus, charge or mass and also all forms of energy as sources are ``built'' from fields and the
conservation of the sources is the consequence of this construction. \\
Based on these assumptions we present two applications, where secondary fields can be generated
from primary fields.
The source-freedom of these secondary fields is an important condition, which prohibits a 
spherical configuration. Thus, spherosymmetrical solutions of the secondary fields
(like monopoles) are excluded.   
In consideration of source-freedom we show that this scalar coupling factor
depends on the value of the primary field energy.
At the point of maximum coupling strength, the strength of the interactions caused from both
primary and secondary fields must be equal.
So the secondary fields can only have either less or equal energy than the primary fields. 
Thus, we have to calculate the values of field energy for the point of maximum coupling strength. \\ 
This paper is organized as follows: \\
In section 2 we see that the consideration of source-free gravitational fields yields to a description
by linear field equations.   
In sec.3 we show that the condition of source-freedom leads to scalar coupling coefficients, which
depend on the square of the field energy of the primary field. 
In sec.4 we present the couplings between primary and secondary fields 
within two applications.
The interactions of primary and secondary fields are discussed in sec.5. 
We determine the value of the maximum coupling strength and their values of field energy in sec.6. 
In sec.7 we discuss the break of the discrete symmetries of parity,
time reversal and charge conjugation. 
Section 8 contains the conclusions. 
\section{Source-free gravitational fields}
\label{sec:2}
The mathematical description of gravitational fields is given by Einstein's
field equations, which contains two contracted forms of the Riemann-Christoffel curvature tensor,
the Ricci tensor and the curvature scalar. \\
Gravitional fields can be compensated locally by the transition to an accelerated
coordinate system but the global field remains, because of the Christoffel symbols are transformed
in a non-linear way. 
The local compensation of the gravitational fields is the result of the equivalence principle. \\
For weak fields we can use a linear approximation by neglecting the self-interaction.
Some components of the Christoffel symbols vanish and the covariant derivatives are reduced
to partial ones so the remaining equations receive a linear structure like
the Maxwell equations in electrodynamics. \\
Further we can derive the Newtonian approximation and equations,
which describe gravitomagnetic fields (p. 197-205 of \cite{flie95}) as a result from the
Thirring-Lense effect \cite{th18}. \\
These linear equations are valid only as an approximation for weak gravitational fields and
constant velocities but not for accelerated coordinate systems. \\ 
To acquire attractive forces among the same kind of charge, we have to set negative signs
at the sources in the linear equations of the vector fields.
Therefore, we receive negative densities of energy and the
transition to accelerated coordinate systems becomes impossible.
This difficulty from negative signs only occurs for a vector theory of gravitation \cite{gu57}. 
The transition to the covariant formalism also has its difficulties especially at the
localization of the gravitational field energy (p. 139 eq.(21.6) of \cite{flie95} and
p. 466-468 of \cite{mis73}). \\
For a vector theory of gravitation we can avoid these difficulties only by homogeneous
field equations, describing fields without any sources.  
In this case the linear field equations are exactly valid not only approximately. \\
The remaining specific components $\hat{\Gamma}$ describing the gravitational vectorfield 
are transformed by linear transformations like a tensor, because the second term vanishes:
\begin{equation}
\hat{\Gamma}'^{\lambda}_{\mu\nu} = \frac{\partial x'^{\lambda}}{\partial x^{\rho}}           
                            \frac{\partial x^{\sigma}}{\partial x'^{\nu}}
                            \frac{\partial x^{\tau}}{\partial x'^{\mu}}
                             \hat{\Gamma}^{\rho}_{\tau\sigma}
\end{equation}
The fact that these components representing the gravitational
vectorfield obey linear transformations causes the vanishing of the following term:
\begin{equation}
\hat{\Gamma}^{\sigma}_{\mu\nu} \hat{\Gamma}^{\rho}_{\sigma\lambda}
 - \hat{\Gamma}^{\sigma}_{\mu\lambda} \hat{\Gamma}^{\rho}_{\sigma\nu} = 0
\end{equation} 
Thus, the Riemann-Christoffel curvature tensor of these components is reduced to the linear form:
\begin{equation}
\hat{R}^{\rho}_{\mu\lambda\nu} = \frac{\partial \hat{\Gamma}^{\rho}_{\mu\nu}}{\partial x^{\lambda}}
                       - \frac{\partial \hat{\Gamma}^{\rho}_{\mu\lambda}}{\partial x^{\nu}}
\end{equation}
Here, the vanishing Riemann-Christoffel curvature tensor is not only a criterion
for accelerated coordinate systems, but here also characterizes the homogeneous
part of a gravitational field. \\
The possibility of different signs of the linear curvature tensor
yields to the fact that the contraction of this curvature tensor has to vanish.
\begin{equation}
\hat{R}_{\mu\nu} = \frac{\partial \hat{\Gamma}^{\rho}_{\mu\nu}}{\partial x^{\rho}} =
 -\frac{1}{2}\Box \hat{g}_{\mu\nu} = 0
\end{equation}  
By the linear field equations follows the additional constraint:
\begin{equation}
\hat{\Gamma}^{\rho}_{\rho\nu} = \frac{1}{\sqrt{\hat{g}}} \frac{\partial\sqrt{\hat{g}}}{\partial x^{\nu}} = 0  
\end{equation}
Therefore, the determinant of the specific metric tensor remains unchanged:
\begin{equation}
\hat{g} = - \det \hat{g}_{\mu\nu} = 1
\end{equation}
Thus, we receive as contributions to the metric tensor $\hat{g}_{\mu\nu}$ only diagonal elements and we always
obtain two invariant remaining spatial components by the source-freedom constraint:
\begin{equation}
\hat{g}_{\mu\nu} = diag \left(\left(1-\frac{2U}{c^{2}}\right),
  \left(\frac{-1}{1-\frac{2U}{c^{2}}}\right), -1, -1 \right)  
\end{equation}
\begin{equation}
U = \left| \vec{g} \vec{x} \right| = \frac{c^{2}}{2} \left| h_{00} \right|
\end{equation}
From the specific form of $\hat{g}_{\mu\nu}$ results no contribution to
the Christoffel symbols.
These specific properties of the gravitational vector field allows the expression
of the specific Christoffel symbols by an antisymmetric field strength tensor and a vector potential 
similar to electrodynamics.
By the transition from covariant derivatives to the partial ones the connections between 
field strength tensors and vector potentials are the same. 
(p. 106 eq.(4.7.2), and p. 108 eq.(4.7.11) of \cite{wei72}). \\ 
(Hence the $g_{\mu\rho}$ stands for any metric tensor.) 
With the four-velocity $u^{\lambda} = \hat{\gamma}(c,\vec{v})$ and $u_{\lambda}u^{\lambda} = c^{2}$,
we receive the following relations for the vector potential and the field strength components:
\begin{equation}
h_{0\mu} = \frac{2}{c} A_{[g]\mu}
\end{equation}
\begin{equation}
g_{\mu\rho} \hat{\Gamma}^{\rho}_{\nu\lambda} \frac{u^{\lambda}}{\hat{\gamma}} 
= \hat{\Gamma}_{\mu,\nu\lambda} \frac{u^{\lambda}}{\hat{\gamma}}
= F_{[g]\mu\nu} = \frac{\partial A_{[g]\nu}}{\partial x^{\mu}} - 
\frac{\partial A_{[g]\mu}}{\partial x^{\nu}}  
\end{equation}
\begin{equation}
\hat{\gamma} = \frac{1}{\sqrt{1-\frac{v^{2}}{c^{2}}}}
\end{equation}
Considering these vector fields described by linear field equations, we arrive at two results
within a common formalism. 
In both cases no self-interaction remains and the equations have an exact
linear structure for all applications: \\
-The field is not a gravitational field, it can be described by the inhomogeneous  
Maxwell equations.
In this case two kind of charges might exist and we receive the equations of electrodynamics. \\
-The field is a gravitational field, therefore it must be source-free and we obtain
with the Lorentz-gauge:
\begin{equation}
\frac{\partial F_{[g]}^{\mu\nu}}{\partial x^{\mu}} = \Box A_{[g]}^{\nu} = \Box h^{0\nu} = 0
\end{equation}
We obtain for the metrical representation by the invariant line element:
\begin{equation}
ds^{2} = \left(1-\frac{2U}{c^{2}}\right)c^{2}dt^{2} + \left(\frac{-1}{1-\frac{2U}{c^{2}}}\right)dx^{2}
 -dy^{2} -dz^{2} + 2 h_{0\mu} dx^{\mu} cdt
\end{equation}
Unlike in general relativity we obtain a local conservation of energy and momentum
of the source-free gravitational fields, because energy and momentum can be localized and these fields
are transformed in a linear way. \\
The gravitational vector field couples to the current density of energy and momentum:
\begin{equation}
 j_{[m]}^{\nu} = \frac{\hat{\gamma}}{c^2} u_{\mu} T^{\mu\nu} 
\end{equation}
For the interaction of the vector field we obtain by the local conservation law: \\
\begin{equation}
- \frac{\partial T^{\mu\nu}}{\partial x^{\mu}} = \hat{\Gamma}^{\nu}_{\mu\lambda} T^{\mu\lambda}  
= F_{[g]}^{\nu\lambda} j_{[m]\lambda}
\end{equation} 
The following effects occur with the source-free gravitational fields as well
as with the gravitational fields in general relativity: \\
-The gravitational red- and blue shift of spectral lines. \\
-The deflection of light. \\
-The time delay of signals and scale contraction. \\
-The precession of a gyroscope. \\ 
Thus, these source-free gravitational fields are able to exert forces on all
masses and forms of energy but cannot be generated by themselves. \\
Especially for the case of dipole fields we see that the conservation of the centre of gravity
also requires a vanishing gravitational dipole moment (p. 975 of \cite{mis73}). \\
Therefore, gravitational dipole radiation (spin-1 graviton or graviphoton)
cannot have sources from mass or energy. 
 \section{The source-free generation of fields}
\label{sec:3}
In all discussed cases the electric or magnetic fields are the only components of the primary field
\begin{math} [e] \end{math}, while \begin{math} [s] \end{math} stands for any secondary field.  
We set for the field strength tensors:
\begin{equation}
F_{[s]}^{\mu\nu} = F_{[e]}^{\mu\nu} * K
\end{equation}
With: \begin{math} F_{[e]}^{\mu\nu}= \end{math} primary field.
      \begin{math} F_{[s]}^{\mu\nu}= \end{math} secondary field.
      \begin{math} K = \end{math} scalar coupling coefficient. \\  
By considering the source-freedom of the secondary field we receive:
\begin{equation}
\frac{\partial F_{[s]}^{\mu\nu}}{\partial x^{\mu}} = 
\frac{\partial F_{[e]}^{\mu\nu}}{\partial x^{\mu}} * K
 + F_{[e]}^{\mu\nu} *\frac{\partial K}{\partial x^{\mu}} = 0   
\end{equation}
For the vector potentials we receive the following conditions:
\begin{equation}
A_{[s]}^{\nu} = A_{[e]}^{\nu} * K \quad \leftrightarrow \quad
A_{[e]}^{\nu} * \frac{\partial K}{\partial x_{\mu}} - 
A_{[e]}^{\mu} * \frac{\partial K}{\partial x_{\nu}} = 0
\end{equation} 
and obtain with the Lorentz-condition for $A_{[e]}^{\nu}$ and $A_{[s]}^{\nu}$ the following terms:
\begin{equation}
\Box A_{[s]}^{\nu} = \Box A_{[e]}^{\nu} * K + A_{[e]}^{\nu} *\Box K = 0
\end{equation}
\begin{equation}
 A_{[e]}^{\nu} * \frac{\partial K}{\partial x^{\nu}} = 0 
\end{equation} 
For the symmetric energy-momentum tensor:
\begin{equation}
T_{[e]}^{\mu\nu} = \frac{1}{\mu_{0}}\left(g^{\mu\rho} F_{[e]\rho\lambda} F_{[e]}^{\lambda\nu}
               + \frac{1}{4}g^{\mu\nu} F_{[e]\rho\lambda} F_{[e]}^{\rho\lambda}\right)
\end{equation}
we receive for the couplings:
\begin{equation}
T_{[s]}^{\mu\nu} = T_{[e]}^{\mu\nu} * K^{2}
\end{equation} 
The divergence-free energy-momentum tensor results from the source-freedom of the secondary field:
\begin{equation}
\frac{\partial T_{[s]}^{\mu\nu}}{\partial x^{\mu}} = 
K * \frac{\partial T_{[e]}^{\mu\nu}}{\partial x^{\mu}} 
+ 2 T_{[e]}^{\mu\nu} * \frac{\partial K}{\partial x^{\mu}} = 0 
\end{equation} 
With the electric current density follows:
\begin{eqnarray}
\mu_{0} j_{[e]}^{\nu} * K + F_{[e]}^{\mu\nu} * \frac{\partial K}{\partial x^{\mu}} &=& 0 \\    
- F_{[e]}^{\nu\lambda} j_{[e]\lambda} + T_{[e]}^{\mu\nu} \frac{2}{K} *
 \frac{\partial K}{\partial x^{\mu}} &=& 0  
\end{eqnarray} 
We can also receive the last two equations by using the covariant divergence, because the 
covariant derivatives and partial derivatives are the same for the scalar coupling coefficients.
The electric current density is calculated from the covariant divergence of the 
electromagnetic field strength tensor. \\
For a stationary consideration we look at the spatial components,
which characterize the local behaviour of the scalar coupling factor 
related to the local variation of the field strength and the energy density:
\begin{eqnarray}
                  (\vec{\nabla}K) * \vec{E} &=& - K * \varrho_{[e]}  \\  
\left((\vec{\nabla}K) \times \vec{B}\right) &=& - K * \vec{j}_{[e]}  \\
                    2 \vec{\nabla}K * E^{2} &=& - K * \varrho_{[e]} * \vec{E} \\   
                    2 \vec{\nabla}K * B^{2} &=& - K * \left(\vec{j}_{[e]} \times \vec{B}\right)
\end{eqnarray}
Here we can see that in the inhomogeneous part of the field
(near the charge and current densities) the coupling coefficient decreases,
which means a local decoupling of the fields.
However, the coupling coefficient reach a local maximum in the homogeneous part of the field. \\
Hitherto, only the local variation was considered. 
However, this gives no information about the absolute 
value of the coupling strength concerning the hole field volume.
For calculating the general variation of the coupling strength, we integrate over the field volume:
\begin{equation}
-\int\limits_{V}\left(\frac{2}{K} T_{[e]}^{\mu\nu} *
 \frac{\partial K}{\partial x^{\mu}}\right) dV =
 \int\limits_{V}\left(\frac{\partial T_{[e]}^{\mu\nu}}{\partial x^{\mu}}\right) dV = 0       
\end{equation}   
We can see after the volume integration that all local variations of the coupling coefficient
as well as of the energy density of the field vanish if the field energy is conserved. 
\begin{equation}
\int\limits_{V}\left(\frac{\partial T_{[e]}^{\mu\nu}}{\partial x^{\mu}}\right) dV =
\int\limits_{\partial V = \sigma} T_{[e]}^{\mu\nu} d\sigma_{\mu}          
\end{equation}
The right-hand side of the previous equation shows the equivalent integration
over the surrounding surface of the volume: \\
\begin{math} \sigma_{\mu}=\left(\frac{\partial V}{\partial x^{\mu}}\right) \end{math} \\
This shows that the general variation of the coupling strength must have a dependency
from the field energy. \\   
The general variation \begin{math} \frac{\partial K}{\partial x^{\mu}} \end{math}
of the coupling strength can be expressed by the energy of the primary field: \\
\begin{math} W_{[e]}^{\mu\nu} = \int\limits_{V}T_{[e]}^{\mu\nu}dV \end{math}   
\begin{equation}
\frac{\partial W_{[e]}^{\mu\nu}}{\partial x^{\mu}} =
\frac{2}{K} W_{[e]}^{\mu\nu} * \frac{\partial K}{\partial x^{\mu}}    
\end{equation}
It is seen from this equation that an increasing field energy provides an increasing coupling strength.   
This equation describes only the relation between the coupling strength $ K $ 
and the primary field energy $ W_{[e]}^{\mu\nu} $ without any interactions
of sources like mass or charge. These interactions are discussed in section 5. \\
We have to include that our scalar coupling coefficient $ K $ must be a Lorentz scalar
and is therefore only dependent from Lorentz invariant values.
We can receive for the fields a similar relativistic energy-momentum equation, from
which a Lorentz scalar is achieved.
This results by the product of the covariant tensor $T^{\mu\nu}$ with
its contravariant counterpart (p. 609 of \cite{jac99}). \\ 
However, in general the volume integration 
$W_{[e]}^{\mu\nu} = \int\limits_{V}T_{[e]}^{\mu\nu}dV$ 
cancels the Lorentz invariance of the resulted scalar.
To restore the Lorentz invariance the volume integration must be restricted to
the rest frame of the field. \\
By the following relation: 
\begin{equation}
P^{\mu}P_{\mu} c^{2} = W_{f}^{2} - p^{2} c^{2} = m_{0}^{2} c^{4} = \left(\frac{1}{2}\right) 
W_{[e]}^{\mu\nu}W_{[e]\mu\nu}
\end{equation}
with the corresponding field values:
\begin{eqnarray}
P^{0} = \frac{W_{f}}{c} &=& \int\limits_{V}\frac{1}{2\mu_{0}c}\left(\frac{E^{2}}{c^{2}} + B^{2}\right)dV \\
P^{i} = \vec{p} &=& \int\limits_{V}\frac{1}{\mu_{0}c^{2}}\left(\vec{E}\times\vec{B}\right)dV
\end{eqnarray}
the Lorentz scalar of the field follows as a rest mass $m_{0}$, which is not a fixed value but
only depends on the field strength and the volume of the field within the rest frame. \\
The possibility of the definition of a field rest frame is connected with the timelike property 
of the 4-vector $P^{\mu}$, which is expressed by the following condition:
\begin{equation}
P^{\mu} P_{\mu} \geq 0
\end{equation}
For fields with spherical symmetry this condition is not strictly uphold.
Also any influence of the secondary fields in a spherical configuration is
cancelled, which prohibits a determination of a Lorentz invariant coupling
coefficient $ K $. \\
The homogeneous part of the field can be considered as a field rest frame
by the fact that there the coupling coefficient $ K $ reaches its local maximum. \\
For the rest frame of the field, expressed by the rest mass $\left( m_{0} \geq 0 \right)$, 
we can have two different cases for the primary field: \\
{\bf Case A} (a field of stationary charge distributions): \\
We only have a field of stationary charges therefore, only a pure electric field exists.
So the rest mass of the primary field is represented by the electric field only. 
\begin{equation}
m_{0}\left(E,V\right) = \frac{1}{c^2} \int\limits_{V}\frac{1}{2\mu_{0}}\left(\frac{E^{2}}{c^{2}}\right)dV  
\end{equation}
{\bf Case B} (a field of stationary current distributions): \\ 
We only have a field of constant currents therefore, only a pure magnetic field exists.
The rest mass of the primary field is represented by the magnetic field only. 
\begin{equation}
m_{0}\left(B,V\right) = \frac{1}{c^2} \int\limits_{V}\frac{1}{2\mu_{0}}\left(B^{2}\right)dV  
\end{equation}
Only a pure electric or a pure magnetic field can be considered as a primary
field, for a generation of a coupled secondary field.
So the value of the scalar coupling coefficient depends on the value of the
Lorentz invariant square of the specific field energy of electric or magnetic field, but never
on both of them within the same case. \\
We see that these applications of couplings
are not possible in the direct way for radiation fields, because real photons have no rest mass.
\begin{equation}
P^{\mu} P_{\mu} = \hbar^{2} \left( \frac{\omega^{2}}{c^{2}} - k^{2} \right) = 0 
\end{equation}
The condition \begin{math}P^{\mu} P_{\mu}\neq 0 \end{math} is only valid for virtual particles like
electric (Coulomb-) photons and magnetic photons, which are off-mass-shell photons.
Thus, any conceivable coupling between electromagnetic dipole radiation and gravitational dipole radiation
is generated by the oscillating primary and secondary fields itself and cannot be generated in the direct
way as in the case between primary and secondary fields. \\ 
We receive for the Lorentz invariant terms of the field:   
\begin{equation}
2\left(W_{[e]}^{\mu\nu}W_{[e]\mu\nu}\right) * \frac{\partial K}{\partial x^{\mu}} =
\frac{1}{2} K * \frac{\partial\left(W_{[e]}^{\mu\nu}W_{[e]\mu\nu}\right)}{\partial x^{\mu}}  
\end{equation}
For this differential equation we need a special nontrivial solution
with the property: \\ 
$ dK \sim K $ and the boundary conditions: \begin{math} K(W_{[e]}=0)=0 \end{math} and 
                                  \begin{math} K(W_{[e]}=W_{0})=K_{0} \end{math}. \\
Hence, no special derivation variable is necessary:
\begin{equation}
4(W_{[e]}^{2}) dK = K d(W_{[e]}^{2}) = K \triangle(W_{[e]}^{2})  
\end{equation}
We integrate after separation: 
\begin{equation}
\int\limits_{K_{0}}^{K(W)} \frac{dK}{K} = \frac{1}{4}\left(\frac{W_{[e]}^{2} - W_{0}^{2}}{W_{[e]}^{2}}\right) 
\end{equation}
and arrive at the final result:
\begin{equation}
K(W_{[e]}) = K_{0} * \exp\left[\frac{1}{4}\left(1-\frac{W_{0}^{2}}{W_{[e]}^{2}}\right)\right]
\end{equation}
\section{The applications of the couplings}
\label{sec:4}
\subsection{Application 1}
The coupling of the source-free gravitational field with the electromagnetic field. \\
We set for the field strength tensors:
\begin{equation}
F_{[g]}^{\mu\nu} = F_{[e]}^{\mu\nu} * K_{[e]}
\end{equation}
written in components of the fields:
\begin{eqnarray}
\vec{E} &*& K_{[e]} = -\vec{g} \\
\vec{B} &*& K_{[e]} = -\vec{\Omega}
\end{eqnarray}
Here the coupling coefficient must have the expression of the specific charge: \\
$K_{[e]} = \left[\left|\frac{e}{m}\right|\right]$ \\ 
\subsection{Application 2}
The coupling of: \\
-a magnetic field with an electric field. \\ 
-an electric field with a magnetic field. \\ 
We set for the field strength tensors:
\begin{equation}
F_{[b]}^{\mu\nu} = F_{[e]}^{\mu\nu} *\frac{K_{[v]}}{c}
\end{equation}
written in components of the fields:
\begin{eqnarray}
\vec{E} &*& \frac{K_{[v]}}{c^{2}} = -\vec{B} \\
\vec{B} &*& K_{[v]} = -\vec{E}
\end{eqnarray}
Here the coupling coefficient must have the expression of the velocity: \\
$K_{[v]} = \left[\left|\vec{v}\right|\right]$ \\
Together with the first application we receive:
\begin{eqnarray}
\left(\vec{E} * \frac{K_{[v]}}{c^{2}}\right) &*& K_{[e]} = \vec{\Omega} \\
\left(\vec{B} * K_{[v]}\right) &*& K_{[e]} = \vec{g}
\end{eqnarray} 
\subsection{The directions of the primary and secondary field components}  
For the linear coupling of the field components we can have two directions 
either the parallel or the opposite parallel direction. \\
From this fact results that the interaction forces of the primary and secondary field can be
either added or subtracted dependending on the sign of the charges. \\
To find the right direction (both directions within the same case are impossible!),
we consider the qualities of the electrically charged particles concerning the interactions within the 
coupled fields. \\
For the electro-gravitational coupling we find that in the case of the coupling along the 
anti-parallel direction, the negative charged particles show additive forces
and positive charged particles show subtractive forces.
To transfer the momentum and energy from primary to the secondary field the forces generated by the two
coupled fields must have the same direction.
This is fulfilled in this case by the negative charged particles.   
We observe this fact in the most materials, the momentum and energy of the electromagnetic current
is indeed mostly carried by negative charged particles especially by the electrons.
Thus, the consideration of the coupling along the anti-parallel direction is justified here. \\
For the coupling between electric and magnetic fields these considerations are more difficult here;
no transfer like in the aforementioned approach occurs in a direct way, because there is no magnetic charge. 
In the coupled electric and magnetic fields the electrically charged particles move along the direction 
of the electric field and their magnetic moments are oriented along the direction of the magnetic field.        
A determined direction of the coupling between these two fields, causes a break of the symmetry between left-handed and right-handed
particles.
An example of a break of the left-right symmetry occurs within the weak
interaction, which prefers the left-handed particles.
According to the preference of the left-handed particles, the consideration of the coupling along the
antiparallel direction is justified. \\
\subsection{The coupling strength of both applications} 
The equations of the coupling coefficients have the same structure for both applications.
The only difference is within the coupling coefficients and the specifc energy values:
\begin{eqnarray}
 K_{[e]} = K_{q} &*& \exp\left[\frac{1}{4}\left(1-\frac{W_{eg}^{2}}{W_{[e]}^{2}}\right)\right] \\
 K_{[v]} = K_{c} &*& \exp\left[\frac{1}{4}\left(1-\frac{W_{eb}^{2}}{W_{[e]}^{2}}\right)\right]
\end{eqnarray}
The strength of the maximum coupling remain constant if the field energy exceeds the values of
\begin{math} W_{eg} \end{math} or \begin{math} W_{eb} \end{math}. \\
The upper boundary of the coupling coefficients \begin{math} K_{q},K_{c}\end{math}
as well as the threshold values of the field energies \begin{math} W_{eg},W_{eb}\end{math}, 
will be calculated in section 6. \\ 
\section{The interactions of the fields}
\label{sec:5}
The source-free secondary fields are characterized as free fields.
Therefore, the interaction terms cannot be calculated directly because of the vanishing
field divergence, which results from the source-freedom constraint.
To calculate these interactions, we determine the Lagrangian of primary fields
and include all interactions, which result from both (primary and secondary) fields. \\
This means that the energy of interaction of the secondary fields is transferred from the primary fields. 
The inclusion of all interactions modifies the current density in the Lagrangian. \\ 
With  \begin{math} c^{2} j_{[m]}^{\nu} = \hat{\gamma} u_{\mu} T^{\mu\nu} \end{math} as a
current density of energy and momentum
and a possible magnetic current density \begin{math}j_{[b]}^{\nu}\end{math}
and the primary current density \begin{math}j_{[e]}^{\nu}\end{math}, 
we receive for the complete Lagrangian:
\begin{equation}
{\cal L} = - \frac{1}{4\mu_{0}} F_{[e]\mu\nu}F_{[e]}^{\mu\nu} - j_{[e,m,b]}^{\nu} A_{[e]\nu}
\end{equation}
with the currents:
\begin{equation}
j_{[e,m,b]}^{\nu} = j_{[e]}^{\nu} + j_{[m]}^{\nu} * K_{[e]} + j_{[b]}^{\nu}*\varepsilon_{0}*K_{[v]}
\end{equation} 
We obtain from the Euler-Lagrange equations of the fields:
\begin{equation}
\frac{\partial F_{[e]}^{\mu\nu}}{\partial x^{\mu}} = \mu_{0} * j_{[e,m,b]}^{\nu}
\end{equation}
and determine for each term describing the specific interaction of the fields and currents:
\begin{eqnarray}
\frac{\partial T_{[E]}^{\mu\nu}}{\partial x^{\mu}} &=& - F_{[e]}^{\nu\lambda} * j_{[e]\lambda} \\
\frac{\partial T_{[G]}^{\mu\nu}}{\partial x^{\mu}} &=& - F_{[e]}^{\nu\lambda} * j_{[m]\lambda} * K_{[e]} \\
\frac{\partial T_{[B]}^{\mu\nu}}{\partial x^{\mu}} &=& - F_{[e]}^{\nu\lambda} * j_{[b]\lambda} *\varepsilon_{0}*K_{[v]}
\end{eqnarray}
We obtain an energy balance equation as a local conservation law by summing up these interaction terms
together with the divergence of the energy density of the primary field and integrating over the volume.
However, the coupling strength must remain at a constant value. \\
As we have shown in sec.3 a volume integration is necessary to include the general variation of the
coupling strength $ K $. Together with the interactions of the fields within each application
we obtain as a local conservation:
\begin{eqnarray}
\frac{\partial T_{[e]}^{\mu\nu}}{\partial x^{\mu}} &=&
\left(\frac{2}{K_{[e]}} T_{[e]}^{\mu\nu} * \frac{\partial K_{[e]}}{\partial x^{\mu}}\right)
+\frac{\partial T_{[E]}^{\mu\nu}}{\partial x^{\mu}}
+\frac{\partial T_{[G]}^{\mu\nu}}{\partial x^{\mu}} \\
\frac{\partial T_{[e]}^{\mu\nu}}{\partial x^{\mu}} &=&
\left(\frac{2}{K_{[v]}} T_{[e]}^{\mu\nu} * \frac{\partial K_{[v]}}{\partial x^{\mu}}\right)
+\frac{\partial T_{[E]}^{\mu\nu}}{\partial x^{\mu}}
+\frac{\partial T_{[B]}^{\mu\nu}}{\partial x^{\mu}}
\end{eqnarray}
The variation of the field energy is on the left-hand side of the equation.
One the right-hand side we have the variation of the coupling strength and the terms of the 
interactions. \\
The primary field strength tensor satisfies the wave equation of the fields:
\begin{equation}
\frac{\partial}{\partial x^{\nu}}\frac{\partial F_{[e]}^{\mu\nu}}{\partial x^{\mu}} = 0
\end{equation}
Thus we receive from the Lagrange equations:
\begin{equation}
\frac{\partial j_{[e,m,b]}^{\nu}}{\partial x^{\nu}} =0
\end{equation}
With the following condition for all currents: 
\begin{equation}
j^{\nu} *\frac{\partial K}{\partial x^{\nu}} = \varrho * u^{\nu} *\frac{\partial K}{\partial x^{\nu}} = 0 \\
\end{equation}
\begin{equation}
\frac{\partial j^{\nu}}{\partial x^{\nu}} = 0
\end{equation}
Therefore follows the local conservation for each current and we obtain generally with each coupling coefficient:
\begin{equation}
u^{\nu} * \frac{\partial K}{\partial x^{\nu}} = 0
\end{equation}
This relation can be verified by the energy dependence relation of the coupling coefficients. \\
(We would obtain the same results by the covariant derivatives, because W and K are scalars.) \\
We set for the derivation of K: 
\begin{equation}
dK = \frac{K}{2} \frac{W_{0}^{2}}{W^{3}} * dW
\end{equation}
From which results:
\begin{equation}
u^{\nu} * \frac{\partial K}{\partial x^{\nu}} = \frac{K}{2} \frac{W_{0}^{2}}{W^{3}} *
\left(u^{\nu} * \frac{\partial W}{\partial x^{\nu}}\right) = 0 
\end{equation}
The fact that $W$ depends only on the field values of the E- or the B- field and on the field volume $V$,
with $W\left(E/B,V\right) = m_{0}\left(E/B,V\right) c^{2}$, yields to the following relation:
\begin{equation}
u^{\nu} * \frac{\partial W}{\partial x^{\nu}} =
u^{\nu} * \left(\frac{\partial W}{\partial\left(E/B\right)_{i}} * \frac{\partial\left(E/B\right)_{i}}{\partial x^{\nu}}
          + \frac{\partial W}{\partial V} * \frac{\partial V}{\partial x^{\nu}}\right) = 0
\end{equation}
where $\left(E/B\right)$ means the E- or B- field components, due to the two cases 
({\bf A} or {\bf B}) for the primary field. \\
From the conservation of the electric current follows the charge conservation and
from the conservation of the mass current follows the conservation of energy and momentum.
Thus, any discussed interaction of primary and secondary fields is only possible with these conserved 
currents. \\ 
The spatial and time components of the four-force of mass and electric charge
including both coupling coefficients are described by the following relations: \\
$\left(m = m_{0} \hat{\gamma}\right)$
\begin{equation}
\left(m*K_{[e]}-q_{e}\right)*\left(\left(\vec{E}+\vec{v}\times\vec{B}\right)-
K_{[v]}*\left(\vec{B}-\frac{\vec{v}}{c^{2}}\times\vec{E}\right)\right)
=\vec{F}
\end{equation}
\begin{equation}
\left(m*K_{[e]}-q_{e}\right)*\left(\vec{E}-K_{[v]}*\vec{B}\right)*\vec{v} = P
\end{equation}
For a concluding remark, we note for the calculation of multipole moments that for the secondary fields
the monopole moment vanishes (because of the source-freedom). Thus, all multipole expansions
begin with the dipole moment as the leading term. 
All higher multipole moments of the secondary fields are composed of dipole moments generated
from those fields. 
\section{The maximum strength of the couplings and their corresponding values of field energies}
\label{sec:6}
The main criterion for the maximum strength of the couplings is the equal strength of both interactions
from the primary field as well as from the secondary field.
This criterion characterizes the maximum values of the coupling coefficients and is justified
by the energy conservation between primary and secondary fields. \\
For the coupling between electromagnetic and gravitational fields we determine the relation
between the quantities of the electric charge and the rest mass of the lightest particles.
The energy of the electric field surrounding that electric charge
can be expressed as the rest mass of the electric charge. \\
In spite of the impossibility to calculate these values from the classical field theory,
this relation of quantities between charge and rest mass can be justified by the quantization
of both values of electric charge and rest mass.  
We have to determine this relation of quantities between the electric charge and
the rest mass from the experimental values. \\
The lightest known electrically charged particels are electrons and the positrons as their
charged couterparts.
We assume that the mass shift between electron and the electron-neutrino as the neutral partner
results exclusively from the electric charge, from which results a connection
between electric charge and rest mass.  
Therefore, we can use their specific charge as a characterization for the upper boundary value
of the aforementioned strength of coupling. \\
For a determined energy value \begin{math} W_{eg} \end{math} we have:
\begin{equation}
 0\leq K_{[e]}\leq K_{q}=\left|\frac{e}{m_{e}}\right|
 \qquad\mbox{for:}\qquad
 0\leq W_{[e]}\leq W_{eg}   
\end{equation}
and:
\begin{equation}
 K_{[e]} = K_{q}=\left|\frac{e}{m_{e}}\right| \qquad\mbox{for:}\qquad
 W_{[e]}\geq W_{eg}    
\end{equation}
with the elementary charge: $q_{e} = e$ and the rest mass $m_{0} = m_{e}$. \\ 
For the coupling between electric and magnetic fields at the other application,
we have the value of the light velocity as the quantity of the upper
boundary value of this strength of coupling. \\
For a determined energy value \begin{math} W_{eb} \end{math} we have:
\begin{equation}
  0\leq K_{[v]}\leq K_{c}= \left| c \right| \qquad\mbox{for:}\qquad
  0\leq W_{[e]}\leq W_{eb}
\end{equation}
and:
\begin{equation}
 K_{[v]} = K_{c}=\left| c \right| \qquad\mbox{for:}\qquad
 W_{[e]}\geq W_{eb} 
\end{equation}
To calculate the values of the field energies
\begin{math} W_{eg} \end{math} and \begin{math} W_{eb} \end{math} we have to compare the strength of
the interactions of the primary and secondary fields.
We set for the relation of these interactions: \\ 
(electromagnetic and gravitational interaction)
\begin{equation}
\frac{W_{[G]}}{W_{[E]}}=\left|\frac{m_{e}*K_{[e]}}{e}\right|=\frac{K_{[e]}}{K_{q}} 
\end{equation}        
(electric and magnetic interaction)
\begin{equation}
\frac{W_{[B]}}{W_{[E]}}=\left|\frac{q_{b}*\varepsilon_{0}*K_{[v]}}{e}\right| 
\end{equation}
For the relation between electric and magnetic fields, we have to define the magnetic charge to compare
these interactions.
The value of this elementary charge $ q_{b} $ is defined by the quantum condition of Dirac \cite{dir31}:
\begin{equation} 
 q_{b} = \Phi_{b} = \frac{2\pi\hbar}{e}  
\end{equation}
This condition also occurs with the Aharonov-Bohm effect \cite{ab59} therefore, there is no evidence
for the existence of magnetic monopoles. \\
We obtain for the comparison of the interactions:
\begin{equation}
\frac{W_{[B]}}{W_{[E]}}=\frac{K_{[v]}}{c} *\frac{1}{2\alpha_{e}} 
\end{equation} 
which differes by the factor of \begin{math} 2\alpha_{e} \end{math}.   
The Sommerfeld constant \begin{math}\alpha_{e}\end{math} is defined by:
\begin{equation}
\alpha_{e} = \frac{e^{2}}{4\pi\varepsilon_{0} \hbar c} 
\end{equation}
So we find for the maximum couplings:
\begin{eqnarray}
 K_{[e]} &=& K_{q} \Longrightarrow W_{[G]} = W_{[E]} \\
 K_{[v]} &=& c \Longrightarrow W_{[B]} = W_{[E]}*\frac{1}{2\alpha_{e}}
\end{eqnarray}
For a description by dimensionless coupling constants similar to the
Sommerfeld constant we can define :
\begin{eqnarray}
 \alpha[K_{[e]}] &=& \alpha_{e} *\frac{K_{[e]}}{K_{q}} \leq \alpha_{e} \\ 
 \alpha[K_{[v]}] &=& \alpha_{e} *\frac{K_{[v]}}{c} \leq \alpha_{e}
\end{eqnarray}
Therefore these coupling coefficients have the property of a running coupling constant.
All these secondary interactions were generated from electromagnetic fields, which are the primary fields
in all these applications and cannot become stronger than the electromagnetic interactions. \\
To calculate the quantities of equal interaction strength we determine the ratio of strength of those
interactions, which have no dependency from any coupling coefficients. 
So we have to calculate the ratio of strength between electromagnetic and gravitational interactions
(electrodynamics and general relativity) at the specific case for electromagnetic dipole fields
and gravitational quadrupole fields. \\
These ratios of strength can be described by the the exchange ratios of spin-1 particles
and spin-2 particles, which can be calculated by the strength of the electromagnetic dipole radiation
and the gravitational quadrupole radiation. \\  
To calculate the ratio of radiations we consider the electron-positron system,
which has a non-vanishing dipole moment.   
(For the strength of the electromagnetic dipole radiation and the strength of the gravitational
quadrupole radiation see for example: p. 208 eq.(67.8) / 216 eq.(70.1) and
p. 423 eq.(110.16) / 424 of \cite{lali92}.) \\
To acquire only the restmass as the source of radiation, we have to calculate nonrelativistically. 
In the electron-positron system we can determine \begin{math} r^{2}\omega^{2} \end{math} 
by the following connection:  
\begin{equation}
r^{2}\omega^{2} = \frac{\alpha_{e}^{2} c^{2}}{4} 
\end{equation}
Then we receive for the ratio of the radiations: 
\begin{equation}
\frac{P_{D}}{P_{Q}}= \frac{5}{12} * \frac{4}{\alpha_{e}^2} * 
\frac{e^{2}}{4\pi\varepsilon_{0}\gamma m_{e}^{2}} =\frac{5}{3\alpha_{e}^2}
 *\frac{W_{[E]}}{W_{[M]}}  
\end{equation}
($\gamma = $G, Newton constant). \\
For an equal exchange ratio of spin-1 particles (dipole radiation) and spin-2 particles \\
(quadrupole radiation) we set \begin{math} P_{D} = P_{Q} \end{math}. \\
Therefore, we can deduce:
\begin{equation}
W_{[E]} = \frac{3\alpha_{e}^2}{5} * W_{[M]}  
\end{equation}
and with \begin{math} W_{[B]} * 2\alpha_{e} = W_{[E]} \end{math}:    
\begin{equation}
W_{[B]} = \frac{3\alpha_{e}}{10} * W_{[M]}  
\end{equation} 
The threshold values are determined by the Planck scale of quantum gravity. \\
The Planck energy is defined as: 
\begin{equation}
W_{P} = \sqrt{\frac{\hbar c^{5}}{\gamma}}  
\end{equation}
To get the energy values we set:
\begin{eqnarray}
    W_{P} &=& W_{[M]} \\
 W_{[eg]} &=& W_{[E]} \\
 W_{[eb]} &=& W_{[B]} 
\end{eqnarray}
From this result the threshold values of field energies:
\begin{eqnarray}
 W_{[eg]} &=& \sqrt{\frac{\hbar c^{5}}{\gamma}} * \frac{3\alpha_{e}^2}{5} \\    
 W_{[eb]} &=& \sqrt{\frac{\hbar c^{5}}{\gamma}} * \frac{3\alpha_{e}}{10}
\end{eqnarray}
These values of field energies have the numerical size of:
\begin{eqnarray}
 W_{[eg]} &=&  62,5*10^{3} J \\
 W_{[eb]} &=&  4.28*10^{6} J \\
    W_{P} &=&  1.95*10^{9} J 
\end{eqnarray}
with: \begin{math} \alpha_{e} = 1/137.035 \end{math} \\ 
We note that: \begin{math} W_{[eb]} > W_{[eg]} \end{math} because of the factor
\begin{math} 2\alpha_{e} \end{math}.
For large distances and classical fields the threshold values of energy 
(\begin{math}W_{eg}\end{math} and \begin{math}W_{eb}\end{math}) for the maximum coupling
strength are different from each other and also different from the the value of the
Planck energy \begin{math}W_{P}\end{math}.\\  
We remember the connection between the scalar coupling coefficients and the threshold values
of field energies:  
\begin{eqnarray}
K_{[e]} &=& K_{q} * \exp\left[\frac{1}{4}\left(1-\frac{W_{eg}^{2}}{W_{[e]}^{2}}\right)\right] \\
K_{[v]} &=& K_{c} * \exp\left[\frac{1}{4}\left(1-\frac{W_{eb}^{2}}{W_{[e]}^{2}}\right)\right]
\end{eqnarray}
These threshold values of the energies \begin{math} W_{eg} \end{math}
and \begin{math} W_{eb} \end{math} are also dependent on the Sommerfeld constant. \\
At small distances this constant \begin{math} \alpha_{e} \end{math}
becomes an effective running coupling constant \begin{math} \alpha_{e}\left(Q^{2}\right) \end{math},
which dependency from the energy (transfer of the particle-momentum) is described by renormalization effects
(see for example p. 594 (eq. 12-127) of \cite{itz80}):
\begin{equation}
\frac{1}{\alpha_{e}\left(Q^{2}\right)} = \frac{1}{\alpha_{e}} + \frac{b_{e}}{2\pi} *
 \ln\left(\frac{\left|Q\right|}{\Lambda}\right)
\end{equation}
(Where $\Lambda$ stands for the fixing point of renormalization, Q for the
particle-momentum transfer and $b_{e} (<0)$ corresponds to the number of the charged fermions.) \\ 
Those effective coupling constants occur in the electroweak and strong interactions, which should
be unified at higher energies \cite{gqw74}. \\
The following equation defines an analogous constant for the gravitational interaction depending
on the energy in a direct way (not by renormalization):
\begin{equation}
\alpha_{g} = \frac{m^{2}\gamma}{\hbar c} =\frac{W^{2}\gamma}{\hbar c^{5}} 
\end{equation}
Thus, we receive at the Planck energy: 
\begin{equation}
\alpha_{g} \longrightarrow 1 \qquad\mbox{for:}\qquad W \longrightarrow W_{P}
\end{equation}
At the Planck scale we should expcect that all interactions of particles have the same strength. 
This justifies the following conditions:
\begin{equation}
\left\{\begin{array}{c} W_{eg} \\ W_{eb} \end{array}\right\} \longrightarrow W_{P}
 \qquad\mbox{for:}\qquad \alpha_{e}\left(Q^{2}\right) \longrightarrow 1
\end{equation}
The case of small distances results that the values $W_{eg}$ and $W_{eb}$
lie near the Planck energy $W_{P}$. \\
The quantization of the gravitational fields (spin-1 graviton)
can be realized in an analogous way as it is possible for the electromagnetic fields
in the QED \cite{gu57,gu50}. 
For spin-2 gravitons the approach of canonical quantization is only possible for linear equations \cite{fie39}.  
However, as we have already noted, this approach is excluded for spin-2 paricles \cite{wei80}. \\
All fields in our discussed approach are part of an Abelian gauge theory
of massless spin-1 particles. \\
We receive for the general Lagrangian in this quantum field theory (analogous to quantum electrodynamics)
with: \begin{math} \tilde{q} = \left(- q_{e} + m * K_{[e]} + q_{b} * \varepsilon_{0}*K_{[v]}\right) \end{math}  
\begin{eqnarray}
 {\cal L_{F}} &=& - \frac{1}{4\mu_{0}} F_{[e]\mu\nu}F_{[e]}^{\mu\nu} \\
{\cal L_{I}} &=& - \tilde{q} A_{[e]\nu} \bar{\psi}\gamma^{\nu}\psi \\
 {\cal L_{D}} &=&  \bar{\psi}\left(i\gamma^{\nu}\partial_{\nu} - m_{0} \frac{c}{\hbar}\right)\psi \\
    {\cal L}  &=& {\cal L_{F}} + {\cal L_{I}} + {\cal L_{D}}
\end{eqnarray}
At the scale of the Planck energy \begin{math}W_{P}\end{math}, the theory of
quantum gravitation must be extended to include a quantization of spin-2
particles, which lies outside from our approach for the present.
\section{The violation of the discrete symmetries} 
\label{sec:7}  
The common interactions of the primary and secondary fields yield
to several consequences concerning the discrete symmetries of parity, time
reversal and charge conjugation.
\begin{center}
\begin{tabular}{| l | c | c | c | c | c | c | c |}
\hline
Symmetry & C & P & T & CP & CT & PT & CPT \\
\cline{1-1}
Interaction&  &   &   &    &    &    &    \\
\hline
E/B      & + & - & - & - & - & + & + \\
\hline
E/G      & - & + & + & - & - & + & - \\
\hline
\end{tabular}
\end{center}
Table: The conservation and violation of discrete symmetries by the connected
interactions with primary and secondary fields of the discussed applications. \\ 
\medskip \\
In the bottom row of the table we find an obvious violation of CPT.
This violation is caused by the violation of the electric charge symmetry.
The PT-symmetry is generally conserved, because no axial vectors, which
conserve CP (like in the weak interaction), occur as sources of the fields. \\ 
To determine the maximum coupling strength between the electromagnetic and gravitational fields,
we have defined in the previous section a relation of quantity between the electric charge and the restmass
of the lightest charged particles (electrons and positrons) independent of their sign of charge.
Therefore, both particles must have the same magnitude of rest mass (p. 158 of \cite{itz80}),
which is also required by the CPT-theorem \cite{sw51}. \\ 
In general, the CPT-symmetry is conserved by every relativistic invariant field theory,
which contains  no coupling of gravitational interactions with other interactions.
Any coupling of gravity with other interactions causes violation of the
charge symmetry, because gravitational charge (mass or energy) can not change its sign
by charge conjugation \cite{thi59}. \\
Therefore, the charge symmetry is violated by the electro-gravitational interaction. \\
For our considerations of the fields result two facts: \\
(i) The electric charges, like mass and energy, are sources of fields, which have a spherical
configuration. \\
(ii) The secondary fields resulting from our coupling approach cannot have such spherical configuration. \\
Thus, it seems that the charge symmetry and the CPT conservation hold for fields with a spherical symmetry
but not for those without spherical symmetry.
\section{Conclusions}
\label{sec:conc}
We investigate the generation of fields from other fields by scalar coupling assuming the priority of the fields 
in comparison with their sources.
This kind of field generation is only possible for fields with a non-spherical configuration.
Therefore, the spherical symmetry of all these fields is generally broken. \\
This procedure allows a generation of gravitational fields, which can be described by linear field equations.
The strengths of the discussed couplings depend on the energy of the electric or magnetic fields.
The maximum coupling strength is determined by the specific electric charge of the lightest charged paricles
(electrons and positrons).   
Thus, we are able to describe interacting electric, magnetic, and gravitational spin-1 fields,
as well as their quantization, within a common formalism.
For the quantum field theory approach we find that the maximum coupling strength is reached at the
Planck energy as one should expect for a quantum field theory of gravitational fields. \\
As we have seen, these couplings of two different fields have important consequences for the discrete
symmetries.
The coupling of gravitational fields with other fields causes an obvious CPT violation. \\
Violations of CPT are also conjectured in string field theory \cite{ell95} and
in the standard model \cite{ell99}. \\
However, it seems that for gravitational fields the CPT theorem is not strictly applicable
because general relativity is not a Poincar\'e-invariant theory \cite{wa80}.
Thus, our investigations yield to the following conclusions: \\
It is known that the CPT-symmetry is conserved for the sources, which always generate spherical fields.
However, in our approach the CPT violation occurs through the coupling between
the electromagnetic and gravitational fields. 
Therefore, there must be a difference between the sources, which are surrounded by fields
with spherical symmetry, and the fields without spherical symmetry, resulting from our approach.           
%%\section*{Acknowledgements:}

\end{document}